\documentclass[letterpaper,english,reprint, aps]{revtex4-1}
\usepackage[T1]{fontenc}
\usepackage[latin9]{inputenc}
\usepackage{color}
\usepackage{mathrsfs}
\usepackage{bm}
\usepackage{amsmath}
\usepackage{amssymb}
\usepackage{graphicx}

\usepackage{siunitx}

\usepackage{appendix}

\makeatletter

\pdfpageheight\paperheight
\pdfpagewidth\paperwidth

\usepackage{cancel}
\DeclareSIUnit\angstrom{\text{\AA}}

\makeatother

\usepackage{babel}
\begin{document}
\preprint{APS/123-QED}
\title{Validation of hydrodynamic and kinetic simulations with a plasma interpenetration ICF hohlraum experiment}
\author{S. E. Anderson}\email{andeste@lanl.gov}\affiliation{Los Alamos National Laboratory, Los Alamos, New Mexico 87545, USA}
\author{L. Chac\'{o}n}\affiliation{Los Alamos National Laboratory, Los Alamos, New Mexico 87545, USA}
\author{A. N. Simakov}\affiliation{Los Alamos National Laboratory, Los Alamos, New Mexico 87545, USA}
\author{B. M. Haines}\affiliation{Los Alamos National Laboratory, Los Alamos, New Mexico 87545, USA}
\author{D. S. Montgomery}\affiliation{Los Alamos National Laboratory, Los Alamos, New Mexico 87545, USA}

\date{\today}

\begin{abstract}
We report on simulations of counter-propagating laser-produced plasmas in an inertial confinement fusion (ICF) hohlraum surrogate, aiming to replicate observations reported by Le Pape \textit{et. al} in recent work \cite{LePape2020}. The conditions of the colliding plasmas are relevant to ICF hohlraums used for indirect-drive ignition, and are obtained both with and without low-density He-gas fill. We compare experimental diagnostics to outputs from simulations using the 1D-2V Vlasov-Fokker-Planck kinetic code iFP and the \texttt{xRAGE} radiation-hydrodynamics code. These include the inferred radial lineouts of inferred ion number fraction and ion and electron temperatures, as well as the reported experimental Thomson-scattering (TS) spectra (compared via synthetic TS diagnostics). We observe that 1D kinetic simulations capture the plasma states reported in the experimental diagnostics quite well. Counter-intuitively, the kinetic simulations capture the gas-fill experiment (expected to be more `hydro-like') better than the vacuum experiment, while the reverse is observed for hydrodynamic simulations. This is attributed to the presence of non-trivial multi-dimensional hydrodynamic effects which are more dominant in the vacuum experiment. These effects  are somewhat inhibited in the gas-fill experiment, permitting quasi-1D kinetic plasma transport to play more of a role in producing plasma interpenetration. Differences between the effects of Maxwellian vs. non-Maxwellian (`full $f$') synthetic TS diagnostics are investigated for the kinetic simulations. We find non-Maxwellian TS spectra differ non-trivially from Maxwellian spectra, which suggests caution may be warranted when applying Maxwellian TS models to infer plasma conditions via backward modeling when kinetic effects may be present.




\end{abstract}
\maketitle
\section{Introduction}
Plasma flow interpenetration is a phenomenon of particular interest inside the hohlraum of inertial confinement fusion (ICF) indirect-drive capsule implosions. Specifically, the coronal plasma produced from the laser-ablated high-Z hohlraum walls (e.g., gold) can interact with the ambient fill gas (if present) as well as the coronal plasma produced from the ablating capsule (e.g., high-density carbon). While hydrodynamic phenomena such as Kelvin-Helmholtz (KH) or Rayleigh-Taylor (RT) instabilities may be present in high-gas-fill hohlraums \cite{Amendt2014,Vandenboomgaerde2016}, transport-driven (i.e., due to finite mean-free-path effects) interpenetration may play a significant role as well, particularly in low-gas-fill and vacuum hohlraum environments where the coronal plasma from the capsule ablator and gold wall may interact \cite{Amendt2015,Hopkins2015,Haines2022,Higginson2022}. Notably, transport-driven interpenetration is particularly difficult to account for in radiation-hydrodynamics (rad-hydro) simulations when long-mean-free-path (kinetic) rather than short-mean-free-path (hydro-like) effects are dominant, as rad-hydro models cannot correctly capture kinetic transport. 


Recently, \citeauthor{LePape2020} presented experimental results from a study at the OMEGA laser facility exploring counter-propagating plasmas in a surrogate ICF indirect-drive hohlraum environment \cite{LePape2020}. The experiments collided laser-driven carbon and gold plasmas with and without a background helium gas fill, and were intended to create a quasi-one-dimensional environment. Their stated goal was to investigate the effects of the presence of the background gas on the plasma interpenetration, varying the environment from one dominated by kinetic effects to one where hydro-like effects dominate.

To date, the state-of-the-art for ICF simulations is radiation-hydrodynamics codes (for example, the \texttt{xRAGE} Eulerian hydro-code \cite{Haines2020a,Gittings2008,Haines2017}). However, in regimes where kinetic effects are present such as the hot, low-density region in a hohlraum interior, kinetic modeling may be necessary to correctly capture the physical trajectory of the system. The iFP Eulerian Vlasov-Fokker-Planck code has been developed to explore kinetic physics in ICF environments \cite{Taitano2018,Taitano2021,Taitano2021b}. Given their purported quasi-one-dimensional nature and hohlraum relevance, the experiments of \citeauthor{LePape2020} are a good candidate for validating and characterizing physics performance of iFP, in comparison to  \texttt{xRAGE}.

In the Le Pape experiments, the primary diagnostic tool was the OMEGA facility's Optical Thomson Scattering (OTS) diagnostic \cite{Froula2006}, wherein a low-power laser is scattered from the plasma and collected as a spectrum. From the spectrum of scattered light, the plasma conditions may be determined. However, while Thomson Scattering (TS) is commonly used to diagnose such experiments, there are a few limitations. Notably, the plasma conditions which produced the scattering may only be inferred by using a backward model of the TS process and the assumed plasma environment. Further, backward models typically assume that the plasma distribution functions are Maxwellian. While this has historically been a necessary limitation (because fitting an arbitrary distribution function is much more difficult than fitting moments of Maxwellian distribution), it is known that non-Maxwellian features can produce non-trivial modifications to resulting TS spectra (see e.g., Refs. \cite{Henchen2018,Henchen2019,Milder2019}). We note that recent work has expanded the capability of inferring non-Maxwellian distributions from arbitrary TS spectra (see e.g., Ref. \cite{Foo2023}).

In this study, we simulate the plasma interpenetration experiments using both  \texttt{xRAGE} (in axisymmetric two-dimensional geometry) \cite{Haines2022,Haines2020a,Gittings2008,Haines2017} and the mature iFP kinetic ICF capsule implosion code (one-dimensional in physical space, two-dimensional in velocity-space). We compare them to the experimental results in Ref. \cite{LePape2020}. We generate synthetic TS spectra both for \texttt{xRAGE} (using a Maxwellian TS forward-model) and iFP (using both Maxwellian and non-Maxwellian ``full-$f$'' TS forward-models). We observe a number of interesting features. Firstly, while the experiment was performed in as close to a one-dimensional (cylindrically-symmetric) manner as possible, there are significant multi-dimensional features present (primarily seeded by the oblique incidence of the lasers on the carbon and gold) that challenge its utility for 1D validation. Nevertheless, we find that quasi-one-dimensional kinetic simulations recover the features of the experimental TS spectra much better than the hydrodynamic simulations. We also find that the hydrodynamic simulations capture the number density profiles of the no-gas-fill experiment (which \emph{ought} to be more kinetic) very well, but not the gas-fill experiment (which ought to be more hydro-like). An analysis suggests this is likely a result of significant multi-dimensional effects producing \emph{apparent} interpenetration. Further, the degree of apparent interpenetration is probably significantly increased by the 300\si{\pico\second} time-averaging of the OTS diagnostic \cite{LePape2020}, as the expanding gold and carbon coronae are moving very quickly, particularly in the vacuum case with no gas fill (estimates from Ref. \cite{LePape2020} of flow speed indicate the carbon/gold fronts could traverse between $\sim$50-200\si{\micro\meter} during the OTS averaging window).  Finally, we compare Maxwellian and non-Maxwellian synthetic TS spectra for the iFP simulations and observe non-trivial differences between them, with the potential to introduce errors into subsequent Maxwellian backward-modeling to infer the moments.

The rest of this paper is organized as follows. Firstly, we give an overview of the plasma interpenetration  experiments described in Ref. \cite{LePape2020}. Next, we summarize the capabilities of the radiation-hydrodynamics code \texttt{xRAGE} and outline the simulation parameters used for the rad-hydro simulations. Following this, we give an overview of the iFP Vlasov-Fokker-Planck code, summarizing its capabilities and the setup and parameters of the kinetic simulations. Then, we briefly discuss the models used for Thomson scattering and how the synthetic TS spectra are created.  The next section presents our simulation results, with comparison to the reported experimental data of Ref. \cite{LePape2020}. Finally, we summarize the work and conclude.

\section{Plasma interpenetration experiment overview}
\begin{figure}
    \centering
       \includegraphics[width=0.75\linewidth]{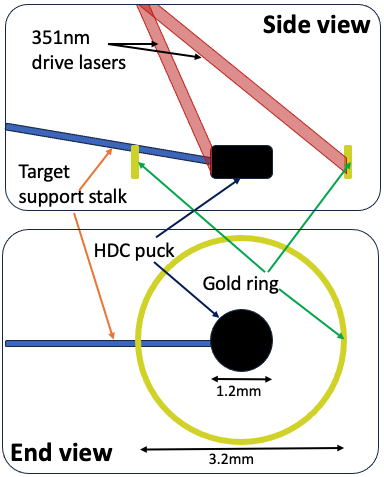}
    \caption{Cartoon of the Le Pape experimental setup (see e.g., Ref. \cite{LePape2020}) showing the view from the side (top) and view from either end (bottom). The relative positions of the inner HDC puck and outer gold ring are shown, with approximate positioning of the target support stalk and drive lasers. Note that the target support stalk is not included in either Vlasov-Fokker-Planck or rad-hydro simuations.  Figure is not to scale though dimensions of HDC puck and inner diameter of the gold band are included.}
    \label{fig:target_cartoon}
\end{figure}

We aim to validate computer simulations against the plasma interpenetration experiments performed by \citeauthor{LePape2020}\cite{LePape2020}. These experiments investigated the effects of gas fill on plasma interpenetration in quasi-one-dimensional  conditions relevant to an ICF indirect-drive (ID) hohlraum. The experimental setup was of a 1200\si{\micro\meter}-diameter high-density-carbon (HDC) puck, 800\si{\micro\meter} in length (which mimics an ICF capsule with HDC ablator), surrounded by a 25\si{\micro\meter} thick gold band, which was 3200\si{\micro\meter}-diameter (outer surface) and also 800\si{\micro\meter} in length (to mimic the ICF hohlraum wall). Two versions of the experiment were performed, one with a vacuum between the HDC and the gold, and another with a low-density (0.15 \si{\milli\gram / \centi\meter^3}) He4 gas bag.  Fig \ref{fig:target_cartoon} gives a basic cartoon of the experimental setup, showing relative sizes and positions of the inner HDC puck and outer gold ring.  Approximate pointing of the drive lasers is also shown, highlighting the oblique incidence angle of the laser illumination.

To drive the ablation and plasma creation, the surface of the HDC and the inner surface of the gold band were both illuminated with 351\si{\nano\meter} laser light. The main laser pulse for both vacuum and gas-fill cases was 300\si{\joule} in a 1\si{\nano\second} flat-top, resulting in 400\si{\tera\watt/\centi\meter^2} laser intensity on the carbon and gold surfaces. In the gas-fill case, the pulse was stepped with 70\si{\joule} in a pre-pulse to burst the gas-bag.

The experiments were diagnosed using Thomson-scattering (TS) with a 40\si{\joule}, 1\si{\nano\second} probe, using 263.23\si{\nano\meter} light, with an approximately 50\si{\micro\meter} focus. To determine the plasma conditions, the usual assumed-Maxwellian TS backward-model was used (see Refs. \cite{LePape2020} and \cite{Froula2011} for details). The conclusion in Ref. \cite{LePape2020} was that the introduction of a low-density helium gas-fill produced a transition from an apparently kinetic interpenetration regime to one dominated by more hydro-like effects due to the inhibiting effect of the background helium fill.

\section{Overview of hydro simulations}
The experimental setup reported by \citeauthor{LePape2020} was replicated in a two-dimensional axisymmetric geometry using \texttt{xRAGE} \cite{Haines2020a,Gittings2008,Haines2017,Haines2022}. \texttt{xRAGE} is the LANL flagship Eulerian multi-physics hydro-code with adaptive mesh refinement (AMR). It includes a laser ray-tracing package with plasma heating via inverse-bremsstrahlung and cross-beam-energy-transfer (CBET) modeling.  Also incorporated are non-local thermal equilibrium (NLTE) radiation transport and emission/opacity modeling as well as a plasma physics module including NLTE variable ionization and ion and electron heat conduction, as well as plasma transport (viscosity).

The hydro simulations were performed using AMR with a minimum/maximum spatial resolution of (1.0 \si{\micro\meter}  / 32.0 \si{\micro\meter}). The numerical laser pulse was set to replicate the pulse shape and energy reported by \citeauthor{LePape2020}\cite{LePape2020}, including the initial foot of the laser pulse intended to break the He4 gas-bag. Further, the laser geometry was arranged as fielded in the OMEGA laser facility. In the vacuum case, the He4 was set to an ultra-low density  gas fill ($\rho_{He4} = 10\si{\micro\gram / cc}$) to mimic vacuum conditions.

\texttt{xRAGE} simulations of gas-fill and vacuum cases were run with and without the hydrodynamic plasma transport model. While there were minor differences in the vacuum case, plasma interpenetration was enhanced in the gas-fill case. For the results presented herein, the plasma transport model was used.

\section{Overview of Vlasov-Fokker-Planck simulations}
The plasma interpenetration experiment was also simulated using the plasma kinetic code iFP \citep{Taitano2018,Taitano2021,Taitano2021b}.  iFP is a 1D-2V (one physical space dimension in $x$, and two velocity-space dimensions, $v_{\parallel} \times v_{\bot}$) Eulerian Vlasov-Fokker-Planck code, which solves the coupled Vlasov-Rosenbluth-Fokker-Planck equations for multiple ion plasma species.
Electrons are modeled as a quasineutral (zero charge density) and ambipolar (zero current density in 1D) fluid with a separate temperature from the ions.
The mesh is adaptive using a mesh-motion technique that minimizes a gradient-based monitor function. The boundary locations are allowed to move according to the same mesh motion strategy, which allows iFP simulations to contract or expand with a given prescribed boundary (e.g., in the case of a spherically symmetric ICF capsule implosion, the interface between pusher and fuel may be tracked). This gives iFP a capability similar to arbitrary Lagrangian-Eulerian (ALE) meshing strategies.  
An extended variable-Coulomb logarithm model was implemented in order to deal with the gold plasma conditions reaching into a moderate coupling regime (due to high density and large effective charge state) \cite{Baalrud2012}. 

For the simulations reported in this study, the physical space geometry was planar, rather than the cylindrical symmetry of the original experiments. This is justified by the domain of interest ranging from radii sufficiently outside the initial carbon surface ($r>600\si{\micro\meter}$ ) to radii sufficiently inside the initial gold surface (r<1600\si{\micro\meter}), such that geometric effects are not expected to dominate the system evolution. 
Based on a simple estimate of a volume of adiabatic expansion of plasma into vacuum, we expect a difference of at most $\sim$30\% in position of the blowoff fronts between (one-dimensional) planar and cylindrical geometry is possible, but without dramatic effects to the overall radial profiles of plasma quantities.
Clearly, there will be differences resulting from multi-dimensional flow that cannot be captured in the one-dimensional case. We expect that these differences would primarily manifest as differences in apparent plasma interpenetration (due to adjacent counter-propagating plasma flows).

iFP's initial and boundary conditions were sourced from the two-dimensional \texttt{xRAGE} simulations. This was done by extracting radial lineouts between the carbon and gold surfaces, which were averaged along the axial direction. 
For the vacuum case, hydro profiles were averaged over $\pm400\si{\micro\meter}$ in the axial direction, the entire axial extent of the gold ring. This was determined as the best choice for this case, due to the significant axial variation in the plasma state observed in the hydro simulations (i.e., `fingering' of the gold and carbon plasmas into the vacuum or ultra-low-density helium), which will be shown later in Sec. \ref{sec:Results}. 
Further, this axial averaging range was chosen to capture these multi-dimensional hydrodynamic `apparent' interpenetration effects in the iFP driving boundary conditions.
For the helium gas-fill case, the axial average was computed over $\pm25\si{\micro\meter}$, much narrower than the vacuum case. This is the stated width of the experimental TS diagnostics (see Ref. \cite{LePape2020}), and -- as will be shown later -- the gas-fill simulations are somewhat less multi-dimensional than the vacuum case.

The initial conditions for the iFP simulations were specified as Maxwellian distributions for the ions based on the profiles of moments (density, velocity, temperature) from the hydro simulations at approximately the start of the main laser pulse. Due to the presence of the pre-pulse used to burst the gas bag the iFP simulations are initialized at $t=0.85 \si{\nano\second}$ in the gas-fill case, while in the vacuum case they are initialized at $t=0.28\si{\nano\second}$.  The iFP boundary conditions on both the gold and carbon boundaries were prescribed by tracking the location of the maximum electron temperature ($T_e$) in the coronal blowoff within $\pm 50\si{\micro\meter}$ of the 50\% critical surface (i.e., where $n_e = 0.5 n_{crit}$, with $n_{crit}$ the density at which the plasma frequency equals the laser frequency). This allows the iFP simulations to capture the effect of the laser drive from the hydro simulations without modeling the laser ray-trace and deposition itself.
The ions are modeled as having a fixed effective average ionization for each species: $Z_{He}=2.0$, $Z_C=5.6$, and $Z_{Au}=32.0$. These ionization levels were chosen from the ionization data in the rad-hydro simulations.

The simulations where run on a 192-cell grid in physical space, with 128x64 cells in $v_{\parallel} \times v_{\bot}$ space. The velocity-space domain is $v_\parallel \in \pm 10 v_{th}$ and $v_\bot \in [0, 10v_{th}]$.  For both vacuum and gas-fill simulations, the average timestep size was 40-50\si{\femto\second}, with a final time around $~1.7\si{\nano\second}$, corresponding to roughly 35,000-40,000 timesteps.

Finally, we note that, for both the iFP and \texttt{xRAGE}  simulations of the ``vacuum'' case, there is in fact an ultra-low-density He gas-fill present (roughly 10\si{\micro\gram / \centi\meter^3}). 
While both \texttt{xRAGE} and iFP can run with an arbitrarily small He-fill density, \texttt{xRAGE} simulations are significantly cheaper with a small but finite He-fill.
For consistency with \texttt{xRAGE}, we retain the same ultra-low He fill density in iFP. 
iFP scoping simulations both with and without the He-fill for the ``vacuum'' case (though retaining an ultra-low "floor" fill for both carbon and gold in both cases) found no appreciable differences. 

\section{Thomson Scattering (TS) and synthetic spectra}
A key element of our comparison to the experimental data is the generation of synthetic TS spectra using the simulated plasma states. The model for the Thomson scattering power spectrum $S(\bm{k},\omega)$ is given by:
\begin{widetext}
\begin{equation}
    S\left(\mathbf{k, \omega}\right) = \frac{2 \pi}{k} \left|1 - \frac{\chi_e}{\epsilon} \right|^2 f_{e,0}\left(\frac{\omega}{k}\right) + \sum_j \frac{2\pi}{k} \frac{Z_j^2 n_j}{n_e} \left|\frac{\chi_e}{\epsilon}\right|^2 f_{j,0} \left(\frac{\omega}{k}\right).
    \label{eq:thomson_spectrum}
\end{equation}
\end{widetext}
In this equation, $\mathbf{k}$ and $\omega$ are the wavevector and frequency, $\epsilon = 1 + \chi_e + \sum_i \chi_i$ is the dielectric function, and $\chi_s$ is the susceptibility for species $s$. The one-dimensional marginal distribution function (i.e., integrated along the perpendicular velocity components) of species $s$ along the wavevector $\mathbf{k}$ is given by $f_{s,0}$. The species $s$ number density is given by $n_s$, with the species effective charge given by $Z_s$.

The susceptibility $\chi_s$ is given by:
\begin{equation}
    \chi_s\left(\bm{k},\omega\right) =  \int_{\infty}^{\infty} d\mathbf{v} \frac{4\pi q_s^2 n_{s,0}}{m_s k^2} \frac{\mathbf{k}\cdot \partial f_{s,0} / \partial \mathbf{v}}{\omega -\mathbf{k}\cdot\mathbf{v}}.\label{eq:susceptibility}
\end{equation}
In the typical TS backward-model for fitting to experimental data, the marginal one-dimensional distribution functions $f_{s,0}$ along the wavevector $\mathbf{k}$ are assumed to be Maxwellian distributions:
\begin{equation}
    f_{s,0} \rightarrow f_{M,s,0} = \frac{n_s}{\pi^{3/2} v_{th,s}^3} e^{\frac{-\left(\mathbf{v}-\mathbf{u}\right)^2}{v_{th,s}^2}}.
\end{equation}
Here, $v_{th,s}=\sqrt{2T_s/m_s}$, with $T_s$ the species temperature and $m_s$ its mass. This assumption greatly simplifies the calculation of the susceptibilities, as Eq. \eqref{eq:susceptibility} may be simplified as in App. D of Ref. \cite{Froula2011}. 

In the case of a non-Maxwellian distribution function, however, the susceptibility integral must be performed numerically, with care being taken to avoid the pole at $\omega = \mathbf{k}\cdot \mathbf{v}$. 
This may be expressed as:
\begin{widetext}
\begin{equation}
    \int_{-\infty}^{\infty} d\mathbf{v} \frac{\partial f_{s,0}/\partial\mathbf{v}}{\omega - \bm{k}\cdot\mathbf{v}} = \int_{-\infty}^{\frac{\omega}{k}-\delta}d\mathbf{v}\frac{\partial f_{s,0}/\partial\mathbf{v}}{\omega - \bm{k}\cdot\mathbf{v}}
    +\int_{\frac{\omega}{k}+\delta}^{\infty}d\mathbf{v}\frac{\partial f_{s,0}/\partial\mathbf{v}}{\omega - \bm{k}\cdot\mathbf{v}}
    +\int_{\frac{\omega}{k}-\delta}^{\frac{\omega}{k}+\delta} d\mathbf{v}\frac{\partial f_{s,0}/\partial\mathbf{v}}{\omega - \bm{k}\cdot\mathbf{v}},
    \label{eq:pole-splitting}
\end{equation}
\end{widetext}
with the integral around the pole expressed as \cite{Henchen2018b}:

\begin{equation}
    \int_{\frac{\omega}{k}-\delta}^{\frac{\omega}{k}+\delta} d\mathbf{v}\frac{\partial f_{s,0}/\partial\mathbf{v}}{\omega - \bm{k}\cdot\mathbf{v}} = 
    -i \pi \left.\frac{\partial f}{\partial \mathbf{v}} \right\vert_{\frac{\omega}{k}}
    +2\delta \left.\frac{\partial^2 f}{\partial \mathbf{v}^2} \right\vert_{\frac{\omega}{k}}.
    \label{eq:pole}
\end{equation}
To compute the synthetic TS spectrum for a given location in space and time, we must specify the scattering angle, which is determined from the experimental TS geometry. In the case of the Le Pape experiments, the scattering angle is 60$^{\circ}$. From the simulations, we have the moments of each of the plasma species (density $n$, velocity $\mathbf{u}$, and temperature $T$) for the Maxwellian TS case, and the full distribution function $f$ in the case of non-Maxwellian TS analysis.

Note that for the non-Maxwellian analysis of the 1D-2V iFP simulations, we technically ought to compute the marginal distribution $f_{s,0}$ along the wavevector $\bm{k}$ (see also the term $\bm{k} \cdot \partial f_{s,0}/ \partial \mathbf{v}$). However, for simplicity, we take the distribution $f_{s,0}$ to be the perpendicularly integrated marginal distribution,
\begin{equation}
    f_{s,0} = 2\pi \int_{0}^{\infty} d\mathrm{v}v_\bot v_\bot f_s(v_{\parallel}, v_{\bot}).
    \label{eq:marginal_perpendicualr}
\end{equation}
While this is not strictly correct, our primary aim is to observe whether accounting for non-Maxwellian features in the TS analysis for this experiment may have a significant impact on the resulting synthetic spectra. As will be shown later, non-Maxwellian features indeed have an impact, which introduces uncertainty in the inferred moments from the standard Maxwellian TS backward model. 

To mimic the experimental TS spectra more accurately, the synthetic spectra were convolved in a number of ways. First, we note that the experimental TS spectra in Ref. \cite{LePape2020} are taken at ``the end on the main laser pulse for a ~300\si{\pico\second} duration''.  Thus, we perform a temporal averaging on our synthetic spectra according to 
\begin{equation}
    \bar{S}(\bm{k},\omega) = \frac{1}{t_1-t_0}\int_{t_0}^{t_1} S(\bm{k}, \omega, t) dt,
\end{equation}
which we approximate on the discrete output via the trapezoidal rule,
\begin{equation}
    \bar{S}(\bm{k}, \omega) = \sum_{k=1}^{N} \frac{S(\bm{k},\omega)_{k-1} + S(\bm{k}, \omega)_{k}}{2(t_N-t_1)} (t_{k} - t_{k-1}).
\end{equation}
Further, to capture effects of uncertaintly in the scattering angle, we average over a range of angles, e.g., $\theta \in [60^{\circ} - \Delta\theta, 60^{\circ} + \Delta \theta]$. Typically $\pm 7 ^{\circ}$ is used here, with a 1$^{\circ}$ resolution. 
To capture the effects of finite resolution in physical space and in the spectrum, we employ a Gaussian blurring kernel convolution in both the spatial and spectral axes. Typical values for characteristic width of the kernels are 50\si{\micro\meter} and 0.5\si{\angstrom}, respectively.

\section{Results and Discussion\label{sec:Results}}
\subsection{Comparison of experimental and synthetic Thomson scattering spectra}
\begin{figure}
    \centering
    \begin{minipage}[b]{0.15\textwidth}
        \includegraphics[width=1.0\linewidth]{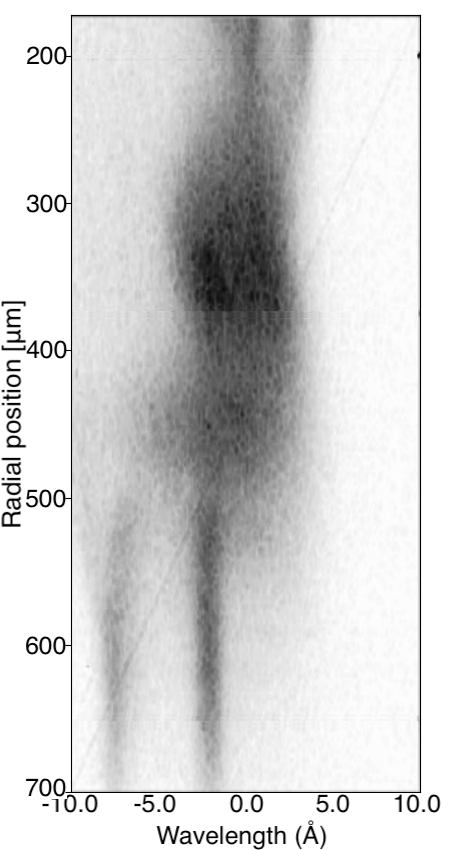}
    \end{minipage}
    \hspace{0.0cm}
    \begin{minipage}[b]{0.15\textwidth}
        \includegraphics[width=1.0\linewidth]{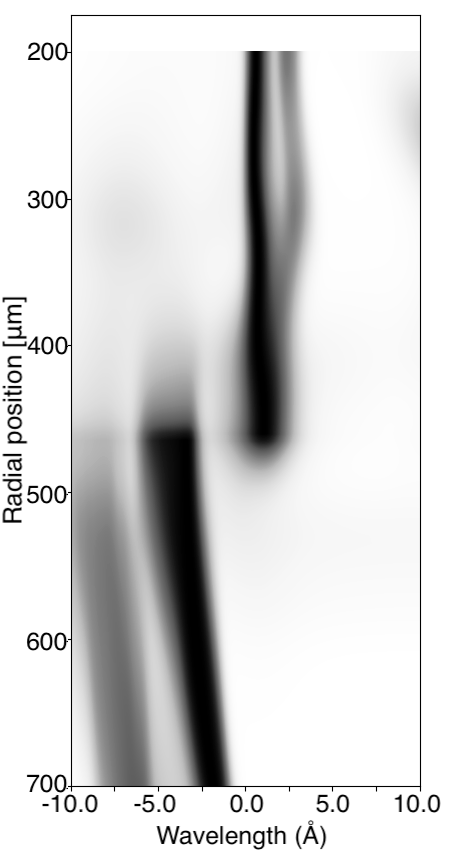}
    \end{minipage}
    \hspace{0.0cm}
    \begin{minipage}[b]{0.15\textwidth}
        \includegraphics[width=1.0\linewidth]{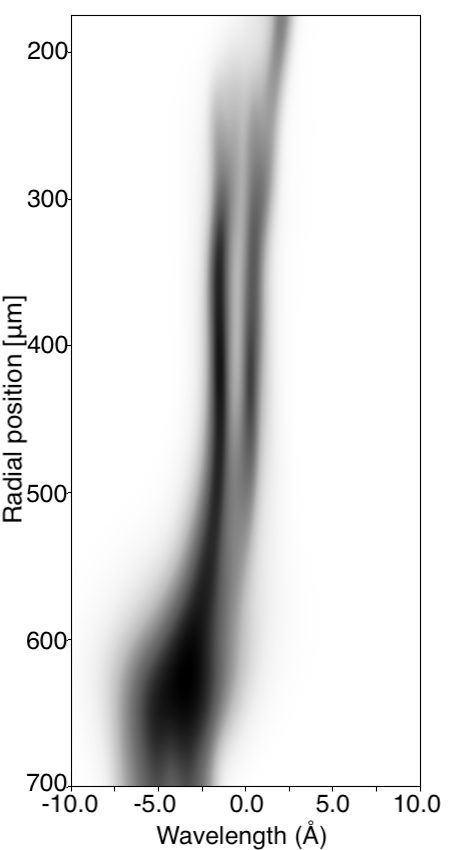}
    \end{minipage}

    \caption{Comparison of experimental OTS spectra (left) to synthetic spectra generated from iFP simulations (center) and \texttt{xRAGE} simulations (right) for the vacuum case. The synthetic spectrum for iFP is computed starting at the end of the main laser pulse ($t=0.88\si{\nano\second}$ after main pulse start) with $300\si{\pico\second}$ temporal averaging with $\sim$12\si{\pico\second} resolution. 
    The \texttt{xRAGE} spectrum is computed using the same axial-averaging as was used to generate the iFP initial and boundary-conditions ($\pm400\si{\micro\meter}$ around axial centerline for this case). }
    \label{fig:TS-vacuum}
\end{figure}

\begin{figure}
    \centering
    \begin{minipage}[b]{0.15\textwidth}
        \includegraphics[width=1.0\linewidth]{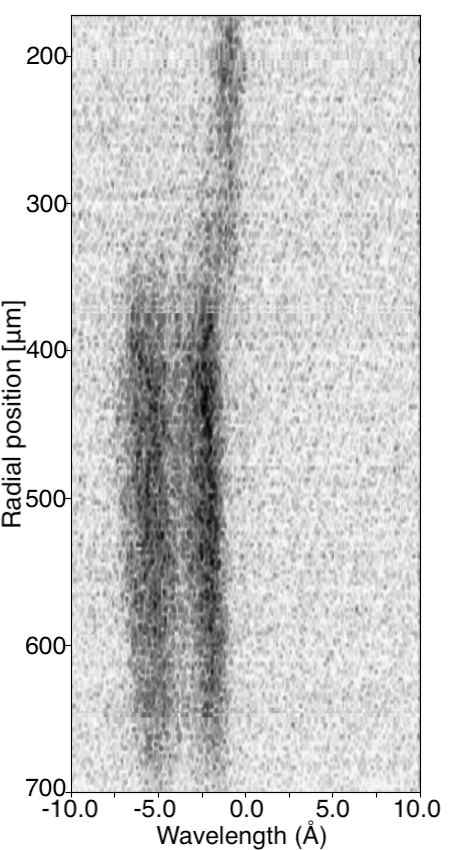}
    \end{minipage}
    \hspace{0.0cm}
    \begin{minipage}[b]{0.15\textwidth}
        \includegraphics[width=1.0\linewidth]{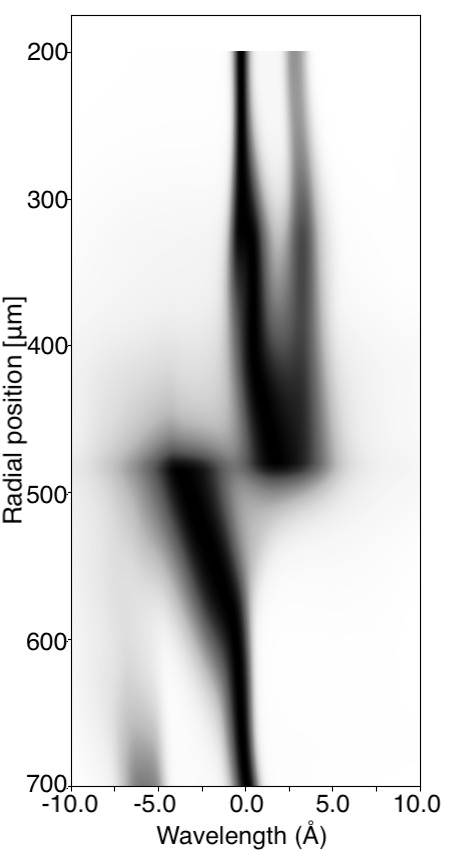}
    \end{minipage}
    \hspace{0.0cm}
    \begin{minipage}[b]{0.15\textwidth}
        \includegraphics[width=1.0\linewidth]{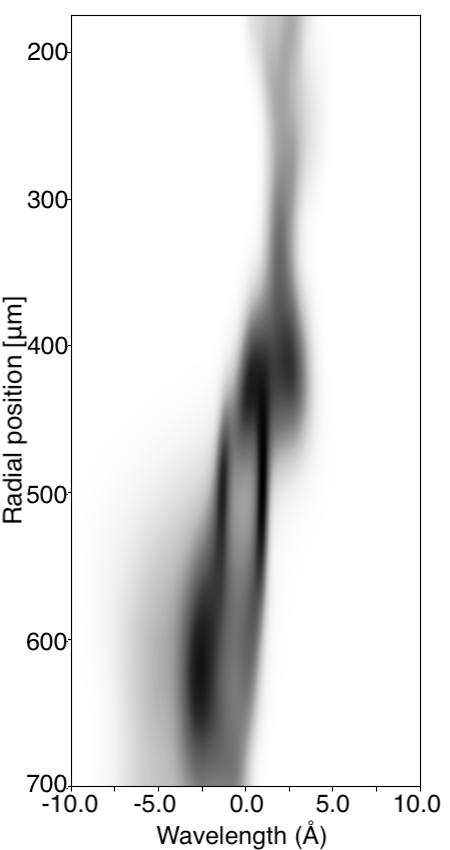}
    \end{minipage}

    \caption{Comparison of experimental OTS spectra (left) to synthetic spectra generated from iFP simulations (center) and \texttt{xRAGE} simulations (right) for the gas-fill case. Synthetic iFP and \texttt{xRAGE} spectra are computed starting at the end of the main laser pulse ($t=1.55\si{\nano\second}$ after pre pulse start) with $300\si{\pico\second}$ temporal averaging $\sim$12\si{\pico\second} resolution. 
    The \texttt{xRAGE} spectrum is computed using the same axial-averaging as was used to generate the iFP initial and boundary-conditions ($\pm25\si{\micro\meter}$). }
    \label{fig:TS-gas}
\end{figure}

Our primary targets for comparison are the experimental OTS spectra (Figs. 2(a) and 2(e) in Ref. \cite{LePape2020}) and synthetic TS spectra generated from the rad-hydro (Maxwellian spectra) and Vlasov-Fokker-Planck (non-Maxwellian spectra) simulations. These comparisons are shown in Figs. \ref{fig:TS-vacuum} and \ref{fig:TS-gas}, showing spectra for the vacuum and gas-fill experiments/simulations, respectively. Here, the position index employs the same convention as in Ref. \cite{LePape2020}, measured from the initial inner surface of the gold, towards the carbon.
We see that the iFP simulations actually produce synthetic TS spectra which agree remarkably well with the qualitative features of the experimental spectra, though much better for the gas-fill case than the vacuum case. For the rad-hydro synthetic spectra, neither gas-fill nor vacuum case matches particularly well. We offer an explanation for this later in the manuscript.

\subsection{Comparison of Maxwellian and non-Maxwellian synthetic TS spectra for iFP simulations}
\begin{figure}
    \centering
    \begin{minipage}[b]{0.23\textwidth}
        \includegraphics[width=1.0\linewidth]{thomson_scattering_t=0.650_to_0.950ns_vacuum.png}
    \end{minipage}
    \begin{minipage}[b]{0.23\textwidth}
    \includegraphics[width=1.0\linewidth]{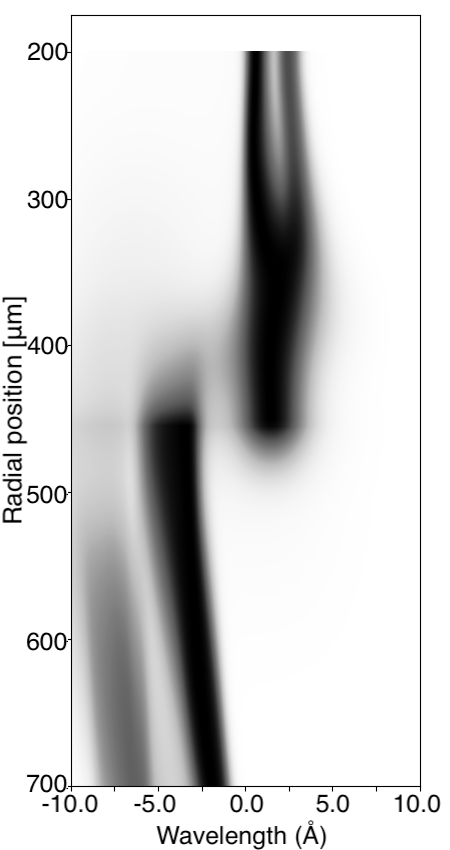}
    \end{minipage}
    \caption{Comparison of non-Maxwellian (left) and Maxwellian (right) synthetic TS spectra for the vacuum case from iFP simulations.}
    \label{fig:TS-vacuum-contour-maxwellian-comparison}
\end{figure}

\begin{figure}
    \centering
    \begin{minipage}[b]{0.23\textwidth}
        \includegraphics[width=1.0\linewidth]{thomson_scattering_t=0.800_to_1.100ns_gas-fill.png}
    \end{minipage}
    \begin{minipage}[b]{0.23\textwidth}
    \includegraphics[width=1.0\linewidth]{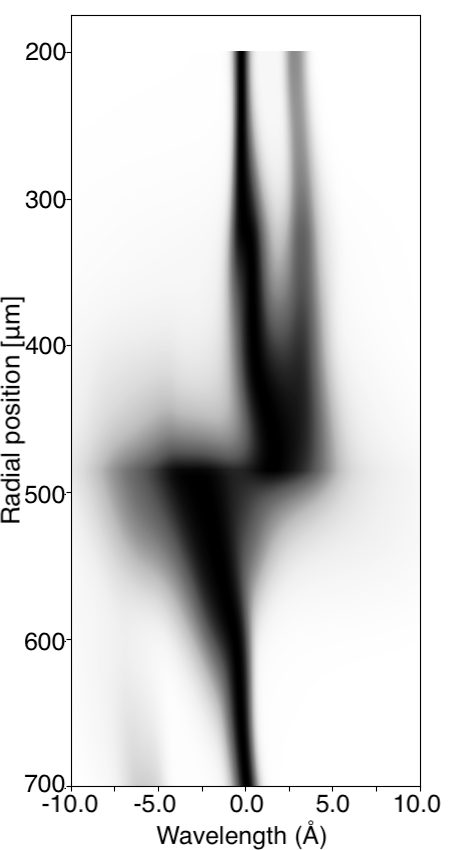}
    \end{minipage}
    \caption{Comparison of non-Maxwellian (left) and Maxwellian (right) synthetic TS spectra for the gas-fill case from iFP simulations.}
    \label{fig:TS-gas-fill-contour-maxwellian-comparison}
    \end{figure}

\begin{figure}
    \centering
    \begin{minipage}[b]{0.23\textwidth}
        \includegraphics[width=1.0\linewidth]{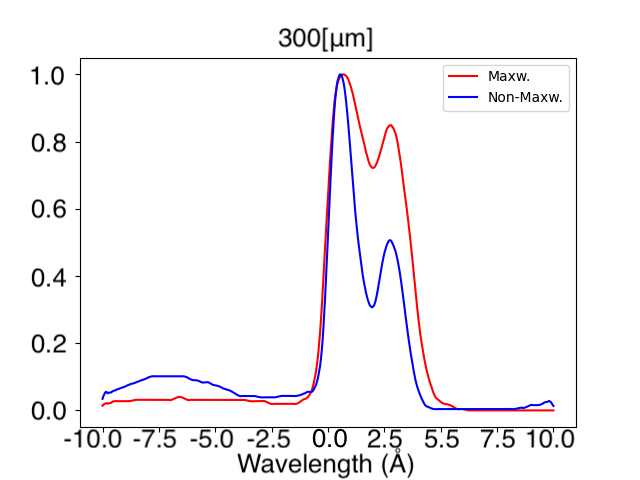}
    \end{minipage}
    \begin{minipage}[b]{0.23\textwidth}
    \includegraphics[width=1.0\linewidth]{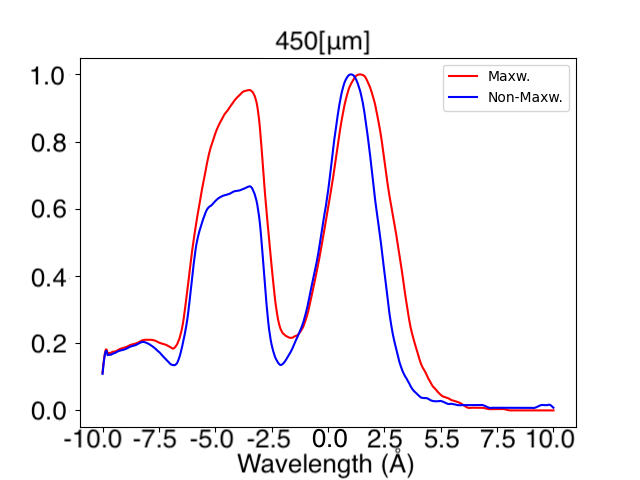}
    \end{minipage}
    \caption{Comparison of non-Maxwellian (blue) and Maxwellian (red) synthetic TS spectra for the vacuum case from iFP simulations, at $300\si{\micro\meter}$ (left) and $450\si{\micro\meter}$ (rigth) from the initial gold interface. Lineouts are taken from the spectra in Fig. \ref{fig:TS-vacuum-contour-maxwellian-comparison}.}
    \label{fig:TS-vacuum-lineout-maxwellian-comparison}
    \end{figure}

\begin{figure}
    \centering
    \begin{minipage}[b]{0.23\textwidth}
        \includegraphics[width=1.0\linewidth]{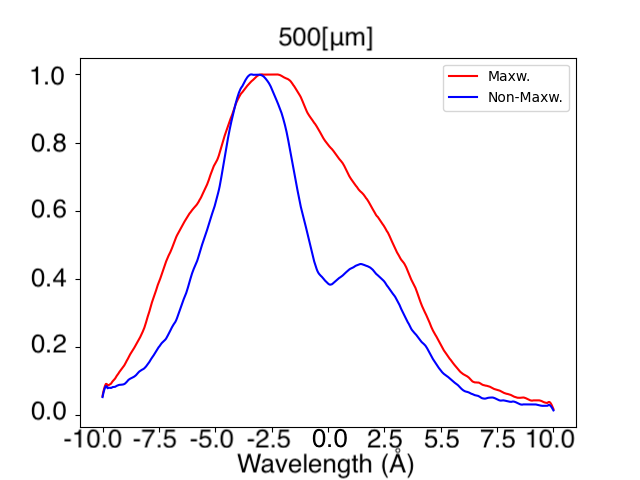}
    \end{minipage}
    \begin{minipage}[b]{0.23\textwidth}
    \includegraphics[width=1.0\linewidth]{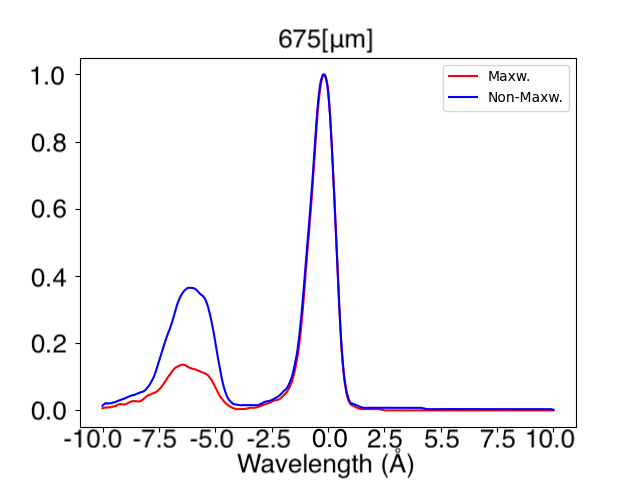}
    \end{minipage}
    \caption{Comparison of non-Maxwellian (blue) and Maxwellian (red) synthetic TS spectra for the gas-fill case from iFP simulations, at $500\si{\micro\meter}$ (left) and $675\si{\micro\meter}$ (rigth) from the initial gold interface. Lineouts are taken from the spectra in Fig. \ref{fig:TS-gas-fill-contour-maxwellian-comparison}.}
    \label{fig:TS-gas-fill-lineout-maxwellian-comparison}
    \end{figure}

Next, we compare the effects of considering a non-Maxwellian model vs. a traditional Maxwellian model for computing synthetic TS spectra. In Figs.  \ref{fig:TS-vacuum-contour-maxwellian-comparison} and \ref{fig:TS-gas-fill-contour-maxwellian-comparison},  we compare Maxwellian and non-Maxwellian spectra for the vacuum and gas-fill iFP simulations, respectively.
We see that, overall, there are noticeable differences, primarily in the center of the domain where long-mean-free-path effects are more dominant. Figs. \ref{fig:TS-gas-fill-lineout-maxwellian-comparison} and \ref{fig:TS-vacuum-lineout-maxwellian-comparison} compare lineouts from Maxwellian and non-Maxwellian synthetic spectra at representative locations in the domain where differences are greatest for the vacuum and gas-fill simulations, respectively. We observe non-trivial differences in the spectral lineouts which may affect the inferred plasma state at these locations using traditional TS backward-modeling. From this, we note that caution is advised when utilizing standard Maxwellian TS models to infer plasma conditions in cases where long-mean-free-path effects are expected to be significant.

\subsection{Comparison of spatial moment profiles}
Finally, we compare the spatial profiles of the plasma moments.  To compare   the \texttt{xRAGE} and iFP moment profiles with the experimentally inferred profiles (see Fig. 3 of Ref. \cite{LePape2020}), we apply similar averaging to that used for the Thomson scattering diagnostics, specifically a 50\si{\micro\meter} Gaussian blurring kernel in space, with 300\si{\pico\second} temporal averaging. In addition, the simulation data is interpolated to radial locations of the given experimental moment profiles.  The timing for the moment profiles is the same as for the prior TS spectra, namely starting at 0.88\si{\nano\second} and 1.55\si{\nano\second} after main laser pulse start for the vacuum and gas-fill cases, respectively. From the previous discussion, we expect fundamental limits in our ability to match experimental data solely based on differences between Maxwellian and non-Maxwellian TS spectra.

\begin{figure}
    \centering
    \begin{minipage}[b]{0.49\textwidth}
        \includegraphics[width=0.9\linewidth]{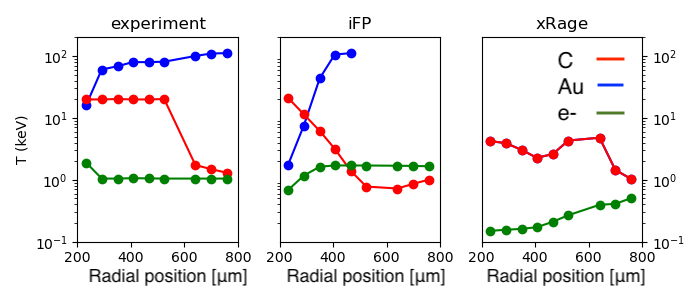}
    \end{minipage}
    \begin{minipage}[b]{0.49\textwidth}
        \includegraphics[width=0.9\linewidth]{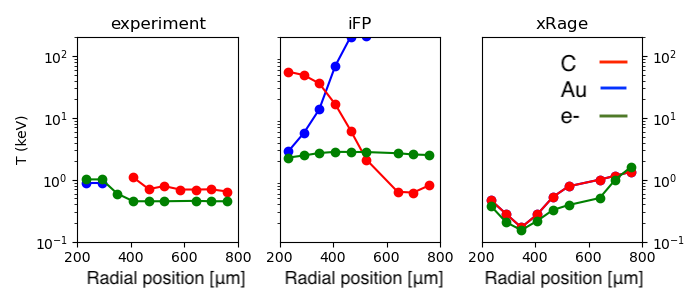}
    \end{minipage}

    \caption{Temperature radial lineouts for iFP and \texttt{xRAGE} simulations, compared to experimental data inferred from a Maxwellian TS model. The plots are oriented using the same convention as Fig. 3 of Ref. \cite{LePape2020}, with the initial gold surface on the left and carbon on the right. The abcissae are spatial positions with respect to the initial inner surface of the gold band. The vacuum case is above, while the gas-fill case is below. We note that the agreement here is much more qualitative in broad trends. Generally, the iFP simulations recover the high gold and carbon temperatures in the vacuum case and the lower electron temperature.}
    \label{fig:temperature}
\end{figure}

For the species temperature profiles, we see in Fig. \ref{fig:temperature} that iFP simulations follow the experimental trends generally better than the rad-hydro simulations (noting \texttt{xRAGE} does not allow for ion temperature separation). We note here that the individual ion species temperature profiles have been reported only in regions where the relative ion number fraction of each species is greater than $10^{-3}$. This is particularly the case with the iFP temperature profiles, where the gold temperature profile is clipped around 500-550\si{\micro\meter} in both cases.

For both cases, iFP is in relatively good agreement with the available inferred experimental data for carbon and gold, capturing the general trends. We note that, for the gas-fill case, the regions where iFP ion temperatures are much higher tend to correspond to regions of lower density, where experimental data are not available. Note also that iFP captures the broadly `flat' electron temperature profiles of the experimental data, and are roughly in line with the electron temperature magnitude. 

We observe that, in general, the iFP  ion and electron temperatures appear overall  higher than the experimentally inferred and the simulated \texttt{xRAGE} temperatures. 
We hypothesize that this is due to iFP not having a radiation package, as radiation in this context would tend to cool electrons and therefore the plasma. For the conditions observed here, we estimate that including radiation with a temperature approximately equal to the electron temperature would reduce ion and electron temperatures by  as much as $\sim$60\%. For example, in the gas-fill case, radiation emission would cool electrons from 2\si{\kilo\electronvolt} to 750\si{\electronvolt}, which is much closer to the experimentally inferred values. Testing this hypothesis will be the subject of future work.

\begin{figure}
    \centering
    \begin{minipage}[b]{0.49\textwidth}
        \includegraphics[width=0.9\linewidth]{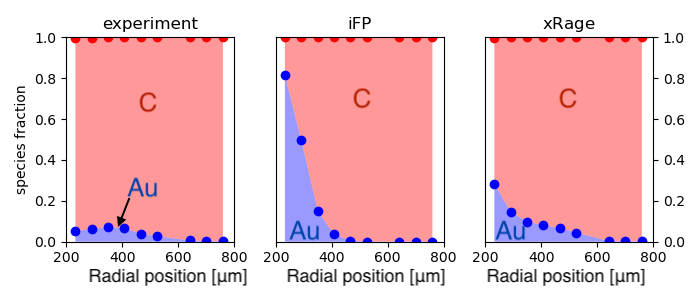}
    \end{minipage}
    \begin{minipage}[b]{0.49\textwidth}
        \includegraphics[width=0.9\linewidth]{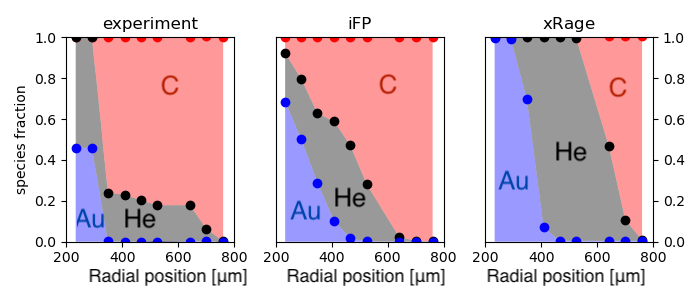}
    \end{minipage}

    \caption{Species relative fraction as a function of radial position for iFP and \texttt{xRAGE} simulations, compared to experimental data inferred from a Maxwellian TS model. The plots are oriented using the same convention as Fig. 3 of Ref. \cite{LePape2020}, with the initial gold surface on the left and carbon on the right. The abcissae are spatial positions with respect to the initial inner surface of the gold band. The vacuum case is above, while the gas-fill case is below.}
    \label{fig:number_fraction}
\end{figure}

Figure \ref{fig:number_fraction} compares the species relative number fractions for the vacuum (above) and gas-fill (below) cases for experiments (left), iFP (center), and \texttt{xRAGE} (right).  Here, carbon is shown in red, gold in blue, and helium (when present) in black. The abcissae are measured relative to the inner surface of the outer gold band.
We observe that, counter-intuitively, \texttt{xRAGE} compares quite favorably to the experimental number fraction profiles for the vacuum case, while iFP does not capture nearly as much apparent interpenetration of the gold and carbon. In the gas-fill case,  \texttt{xRAGE} shows much sharper boundaries between the gold, carbon, and helium regions, indicating very little apparent interpenetration. In this case iFP shows a much more significant degree of interpenetration of the helium into both carbon and gold, and is much closer to the experimentally inferred number fraction profiles.

\begin{figure}
    \centering
    \begin{minipage}[b]{0.8\linewidth}
    \centering
        \includegraphics[width=3in]{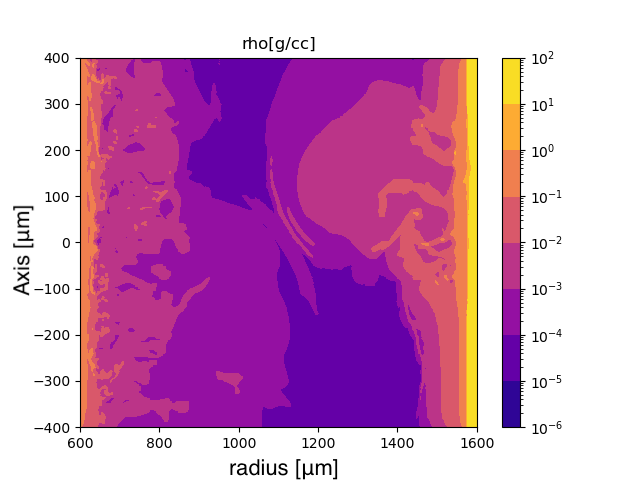}
    \end{minipage}
    \vspace{0.0cm}
    \begin{minipage}[b]{0.8\linewidth}
    \centering
        \includegraphics[width=3in]{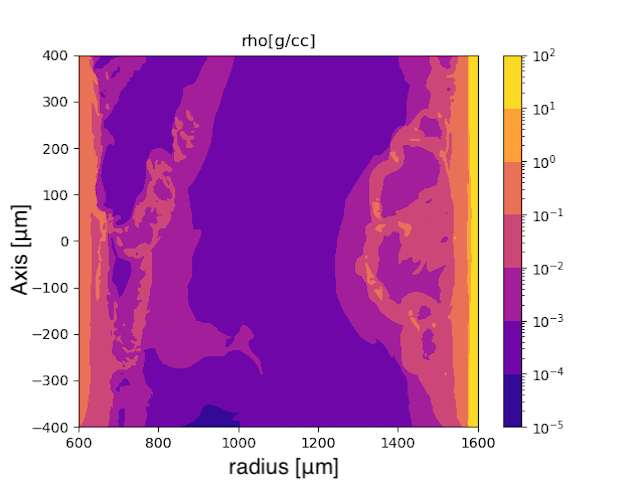}
    \end{minipage}    
    
    \caption{Mass density contour snapshots for 2D axisymmetric \texttt{xRAGE} simulations. The vacuum case is shown on top, and the gas-fill case on bottom. Snapshots are taken at the end of main laser pulse ($1.55\si{\nano\second}$ for gas-fill and $0.88\si{\nano\second}$ for vacuum). Note the presence of significant multi-dimensional features at gold (right-hand side) and carbon (left-hand side) fronts, particularly for the vacuum case.}
    \label{fig:xRAGE_2D}
\end{figure}

Note that there is a general trend of \texttt{xRAGE} appearing to have greater species interpenetration than iFP in the vacuum case while this reverses for the gas-fill case. This is not due to interpenetration of the plasma species, but rather to multi-dimensional hydrodynamic effects, `fingers' of carbon and gold plasmas passing \textit{past} each other, producing \textit{apparent} interpenetration. In fact, these sorts of features can be observed in the two-dimensional axisymmetric \texttt{xRAGE} simulations in Fig. \ref{fig:xRAGE_2D}, where the vacuum case (top) exhibits much more multi-dimensionality than the gas-fill case (bottom), which is inhibited by the presence of the helium gas-fill. iFP, which cannot capture these multidimensional effects, but \textit{can} capture plasma interpenetration due to kinetic plasma transport, shows less interpenetration than \texttt{xRAGE} in the vacuum case, and \textit{more} interpenetration than \texttt{xRAGE} in the gas-fill case.

\begin{figure}
    \centering
    \includegraphics[width=1.5in]{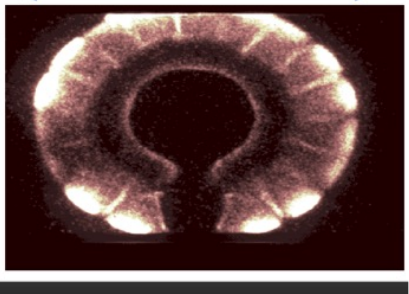}
    \caption{End-on self-emission radiograph from vacuum experiment showing presence of azimuthal features in blowoff~\cite{LePape2019}.}
    \label{fig:azimutha_radiograph}
\end{figure}

Finally, we also note in Fig. \ref{fig:number_fraction}-top (vacuum case) that the interface between the carbon and gold is offset to the right (radially inward) in the experimental and \texttt{xRAGE} number fraction profiles relative to the apparent interface for iFP.  In general, this is the sort of trend we expect from the use of planar geometry in the iFP simulations relative to the cylindrical geometry of the experiment and \texttt{xRAGE} simulations. That is, we expect the blowoff from the gold surface will be somewhat faster in the cylindrical geometry than in planar (converging from a larger to a smaller radius), while we expect the opposite from the carbon surface blowoff (expanding from smaller to larger radius).   Taking the approximate iFP gold-carbon interface as $\sim$400\si{\micro\meter}, and the experimental/\texttt{xRAGE} interface to be $\sim$500-600\si{\micro\meter}, we observe a difference of $\sim$20-30\%, which is in good agreement with the previously stated estimate. The differences in the interface locations are less extreme in the gas-fill case, being inhibited significantly by the presence of the helium.

\section{Summary and Conclusions}
In this study, we have presented a validation exercise with the laser-driven plasma-interpenetration hohlraum surrogate experiments described in Ref. \cite{LePape2020} using both kinetic Vlasov-Fokker-Planck (iFP) and multi-physics radiation-hydrodynamics (\texttt{xRAGE}) models.

The validation exercise is {\em a priori} hampered by the fact that these experiments remain inherently three-dimensional, despite the significant efforts of the authors of Ref. \cite{LePape2020} to minimize high-dimensional effects. This is clear from the two-dimensional rad-hydro simulations depicted in Fig. \ref{fig:xRAGE_2D}, as well as from self-emission imaging from the experiments (see Fig. \ref{fig:azimutha_radiograph}). Nevertheless, useful understanding has been gained from this exercise, which we have reported here.

We have observed that, in general, synthetic TS spectra from the Vlasov-Fokker-Planck model agree much better with the experimental data than those from the radiation-hydrodynamics model. We have also observed non-trivial differences in synthetic TS spectra generated from Vlasov-Fokker-Planck simulations using fully kinetic (`high-fidelity') forward models vs. the usual Maxwellian ansatz. This suggests caution is warranted when interpreting inferred plasma quantities from Maxwellian backward models of experimental TS spectra when kinetic effects may be present.

When comparing experimentally inferred moment profiles to those from simulations, the picture is murkier. \texttt{xRAGE}'s radial species-fraction profiles compare better (counter-intuitively) for the vacuum experiment (which is more kinetic) than the He gas-fill experiment (more `hydro-like'), while failing to predict temperature lineouts altogether. Conversely, iFP achieves better agreement with the number fraction profiles of the gas-fill experiment than the vacuum experiment, while agreeing reasonably closely with experimentally inferred temperature profiles on both experiments. The apparent contradiction in the species-fraction comparison resolves when accounting for the extreme multidimensional nature of the vacuum experiment (due to `fingering' of carbon and gold plasmas passing \textit{past} each other, producing \textit{apparent} interpenetration) vs. the relatively quiet gas-fill experiment, in which the He-fill inhibits such behavior.

Overall, despite iFP's limitations (1D planar, no radiation package, no laser package), we conclude that iFP compares more favorably with key signatures of the experimental data than \texttt{xRAGE}, despite the latter featuring more accurate geometry (two-dimensional cylindrical symmetry), a consistent laser-drive package, radiation-diffusion modeling, as well as tabular dynamic ionization modeling. This suggests that high-fidelity modeling of these experiments in multiple dimensions including kinetic plasma and radiation transport effects is warranted to explain them fully.

\acknowledgements{
The authors thank Sebastien Le Pape, William Taitano, Guangye Chen and Adam Stanier for fruitful discussions. This work was
supported by the US Department of Energy through the Los
Alamos National Laboratory. Los Alamos National Laboratory is operated by Triad National Security, LLC, for the National Nuclear Security Administration of US Department of
Energy (Contract No. 89233218CNA000001). Research presented in this article was supported by the Laboratory Directed
Research and Development program of Los Alamos National
Laboratory under project number 20210063DR. 
This research used resources provided by the Los Alamos National Laboratory Institutional Computing Program.
}

\bibliographystyle{aapmrev4-2}
\bibliography{library,library_2}

\begin{thebibliography}{22}%
\makeatletter
\providecommand \@ifxundefined [1]{%
 \@ifx{#1\undefined}
}%
\providecommand \@ifnum [1]{%
 \ifnum #1\expandafter \@firstoftwo
 \else \expandafter \@secondoftwo
 \fi
}%
\providecommand \@ifx [1]{%
 \ifx #1\expandafter \@firstoftwo
 \else \expandafter \@secondoftwo
 \fi
}%
\providecommand \natexlab [1]{#1}%
\providecommand \enquote  [1]{``#1''}%
\providecommand \bibnamefont  [1]{#1}%
\providecommand \bibfnamefont [1]{#1}%
\providecommand \citenamefont [1]{#1}%
\providecommand \href@noop [0]{\@secondoftwo}%
\providecommand \href [0]{\begingroup \@sanitize@url \@href}%
\providecommand \@href[1]{\@@startlink{#1}\@@href}%
\providecommand \@@href[1]{\endgroup#1\@@endlink}%
\providecommand \@sanitize@url [0]{\catcode `\\12\catcode `\$12\catcode
  `\&12\catcode `\#12\catcode `\^12\catcode `\_12\catcode `\%12\relax}%
\providecommand \@@startlink[1]{}%
\providecommand \@@endlink[0]{}%
\providecommand \url  [0]{\begingroup\@sanitize@url \@url }%
\providecommand \@url [1]{\endgroup\@href {#1}{\urlprefix }}%
\providecommand \urlprefix  [0]{URL }%
\providecommand \Eprint [0]{\href }%
\providecommand \doibase [0]{https://doi.org/}%
\providecommand \selectlanguage [0]{\@gobble}%
\providecommand \bibinfo  [0]{\@secondoftwo}%
\providecommand \bibfield  [0]{\@secondoftwo}%
\providecommand \translation [1]{[#1]}%
\providecommand \BibitemOpen [0]{}%
\providecommand \bibitemStop [0]{}%
\providecommand \bibitemNoStop [0]{.\EOS\space}%
\providecommand \EOS [0]{\spacefactor3000\relax}%
\providecommand \BibitemShut  [1]{\csname bibitem#1\endcsname}%
\let\auto@bib@innerbib\@empty
\bibitem [{\citenamefont {{Le Pape}}\ \emph {et~al.}(2020)\citenamefont {{Le
  Pape}}, \citenamefont {Divol}, \citenamefont {Huser}, \citenamefont {Katz},
  \citenamefont {Kemp}, \citenamefont {Ross}, \citenamefont {Wallace},\ and\
  \citenamefont {Wilks}}]{LePape2020}%
  \BibitemOpen
  \bibfield  {author} {\bibinfo {author} {\bibfnamefont {S.}~\bibnamefont {{Le
  Pape}}}, \bibinfo {author} {\bibfnamefont {L.}~\bibnamefont {Divol}},
  \bibinfo {author} {\bibfnamefont {G.}~\bibnamefont {Huser}}, \bibinfo
  {author} {\bibfnamefont {J.}~\bibnamefont {Katz}}, \bibinfo {author}
  {\bibfnamefont {A.}~\bibnamefont {Kemp}}, \bibinfo {author} {\bibfnamefont
  {J.~S.}\ \bibnamefont {Ross}}, \bibinfo {author} {\bibfnamefont
  {R.}~\bibnamefont {Wallace}},\ and\ \bibinfo {author} {\bibfnamefont
  {S.}~\bibnamefont {Wilks}},\ }\href
  {https://doi.org/10.1103/PhysRevLett.124.025003} {\bibfield  {journal}
  {\bibinfo  {journal} {Physical Review Letters}\ }\textbf {\bibinfo {volume}
  {124}},\ \bibinfo {pages} {25003} (\bibinfo {year} {2020})}\BibitemShut
  {NoStop}%
\bibitem [{\citenamefont {Amendt}\ \emph {et~al.}(2014)\citenamefont {Amendt},
  \citenamefont {Ross}, \citenamefont {Milovich}, \citenamefont {Schneider},
  \citenamefont {Storm}, \citenamefont {Callahan}, \citenamefont {Hinkel},
  \citenamefont {Lasinski}, \citenamefont {Meeker}, \citenamefont {Michel},
  \citenamefont {Moody},\ and\ \citenamefont {Strozzi}}]{Amendt2014}%
  \BibitemOpen
  \bibfield  {author} {\bibinfo {author} {\bibfnamefont {P.}~\bibnamefont
  {Amendt}}, \bibinfo {author} {\bibfnamefont {J.~S.}\ \bibnamefont {Ross}},
  \bibinfo {author} {\bibfnamefont {J.~L.}\ \bibnamefont {Milovich}}, \bibinfo
  {author} {\bibfnamefont {M.}~\bibnamefont {Schneider}}, \bibinfo {author}
  {\bibfnamefont {E.}~\bibnamefont {Storm}}, \bibinfo {author} {\bibfnamefont
  {D.~A.}\ \bibnamefont {Callahan}}, \bibinfo {author} {\bibfnamefont
  {D.}~\bibnamefont {Hinkel}}, \bibinfo {author} {\bibfnamefont
  {B.}~\bibnamefont {Lasinski}}, \bibinfo {author} {\bibfnamefont
  {D.}~\bibnamefont {Meeker}}, \bibinfo {author} {\bibfnamefont
  {P.}~\bibnamefont {Michel}}, \bibinfo {author} {\bibfnamefont
  {J.}~\bibnamefont {Moody}},\ and\ \bibinfo {author} {\bibfnamefont
  {D.}~\bibnamefont {Strozzi}},\ }\href@noop {} {\bibfield  {journal} {\bibinfo
   {journal} {Physics of Plasmas}\ }\textbf {\bibinfo {volume} {21}},\ \bibinfo
  {pages} {112703} (\bibinfo {year} {2014})}\BibitemShut {NoStop}%
\bibitem [{\citenamefont {Vandenboomgaerde}, \citenamefont {Bonnefille},\ and\
  \citenamefont {Gauthier}(2016)}]{Vandenboomgaerde2016}%
  \BibitemOpen
  \bibfield  {author} {\bibinfo {author} {\bibfnamefont {M.}~\bibnamefont
  {Vandenboomgaerde}}, \bibinfo {author} {\bibfnamefont {M.}~\bibnamefont
  {Bonnefille}},\ and\ \bibinfo {author} {\bibfnamefont {P.}~\bibnamefont
  {Gauthier}},\ }\href@noop {} {\bibfield  {journal} {\bibinfo  {journal}
  {Physics of Plasmas}\ }\textbf {\bibinfo {volume} {23}},\ \bibinfo {pages}
  {052704} (\bibinfo {year} {2016})}\BibitemShut {NoStop}%
\bibitem [{\citenamefont {Amendt}, \citenamefont {Ho},\ and\ \citenamefont
  {Jones}(2015)}]{Amendt2015}%
  \BibitemOpen
  \bibfield  {author} {\bibinfo {author} {\bibfnamefont {P.}~\bibnamefont
  {Amendt}}, \bibinfo {author} {\bibfnamefont {D.~D.}\ \bibnamefont {Ho}},\
  and\ \bibinfo {author} {\bibfnamefont {O.~S.}\ \bibnamefont {Jones}},\
  }\href@noop {} {\bibfield  {journal} {\bibinfo  {journal} {Physics of
  Plasmas}\ }\textbf {\bibinfo {volume} {22}},\ \bibinfo {pages} {040703}
  (\bibinfo {year} {2015})}\BibitemShut {NoStop}%
\bibitem [{\citenamefont {Berzak~Hopkins}\ \emph {et~al.}(2015)\citenamefont
  {Berzak~Hopkins}, \citenamefont {Le~Pape}, \citenamefont {Divol},
  \citenamefont {Meezan}, \citenamefont {Mackinnon}, \citenamefont {Ho},
  \citenamefont {Jones}, \citenamefont {Khan}, \citenamefont {Milovich},
  \citenamefont {Ross}, \citenamefont {Amendt}, \citenamefont {Casey},
  \citenamefont {Celliers}, \citenamefont {Peterson}, \citenamefont {Ralph},\
  and\ \citenamefont {Rygg}}]{Hopkins2015}%
  \BibitemOpen
  \bibfield  {author} {\bibinfo {author} {\bibfnamefont {L.~F.}\ \bibnamefont
  {Berzak~Hopkins}}, \bibinfo {author} {\bibfnamefont {S.}~\bibnamefont
  {Le~Pape}}, \bibinfo {author} {\bibfnamefont {L.}~\bibnamefont {Divol}},
  \bibinfo {author} {\bibfnamefont {N.~B.}\ \bibnamefont {Meezan}}, \bibinfo
  {author} {\bibfnamefont {A.~J.}\ \bibnamefont {Mackinnon}}, \bibinfo {author}
  {\bibfnamefont {D.~D.}\ \bibnamefont {Ho}}, \bibinfo {author} {\bibfnamefont
  {O.~S.}\ \bibnamefont {Jones}}, \bibinfo {author} {\bibfnamefont
  {S.}~\bibnamefont {Khan}}, \bibinfo {author} {\bibfnamefont {J.~L.}\
  \bibnamefont {Milovich}}, \bibinfo {author} {\bibfnamefont {J.~S.}\
  \bibnamefont {Ross}}, \bibinfo {author} {\bibfnamefont {P.}~\bibnamefont
  {Amendt}}, \bibinfo {author} {\bibfnamefont {P.}~\bibnamefont {Casey}},
  \bibinfo {author} {\bibfnamefont {P.~M.}\ \bibnamefont {Celliers}}, \bibinfo
  {author} {\bibfnamefont {J.~L.}\ \bibnamefont {Peterson}}, \bibinfo {author}
  {\bibfnamefont {J.}~\bibnamefont {Ralph}},\ and\ \bibinfo {author}
  {\bibfnamefont {J.~R.}\ \bibnamefont {Rygg}},\ }\href@noop {} {\bibfield
  {journal} {\bibinfo  {journal} {Physics of Plasmas}\ }\textbf {\bibinfo
  {volume} {22}},\ \bibinfo {pages} {056318} (\bibinfo {year}
  {2015})}\BibitemShut {NoStop}%
\bibitem [{\citenamefont {Haines}\ \emph {et~al.}(2022)\citenamefont {Haines},
  \citenamefont {Keller}, \citenamefont {Long}, \citenamefont {KcKay},
  \citenamefont {Medin}, \citenamefont {Park}, \citenamefont {Rauenzhan},
  \citenamefont {Scott}, \citenamefont {Anderson}, \citenamefont {Collins},
  \citenamefont {Green}, \citenamefont {Marozas}, \citenamefont {McKenty},
  \citenamefont {Peterson}, \citenamefont {Vold}, \citenamefont {Di~Stefano},
  \citenamefont {Lester}, \citenamefont {Sauppe}, \citenamefont {Stark},\ and\
  \citenamefont {Velechovsky}}]{Haines2022}%
  \BibitemOpen
  \bibfield  {author} {\bibinfo {author} {\bibfnamefont {B.~M.}\ \bibnamefont
  {Haines}}, \bibinfo {author} {\bibfnamefont {D.~E.}\ \bibnamefont {Keller}},
  \bibinfo {author} {\bibfnamefont {K.~P.}\ \bibnamefont {Long}}, \bibinfo
  {author} {\bibfnamefont {M.~D.}\ \bibnamefont {KcKay}, \bibfnamefont {Jr.}},
  \bibinfo {author} {\bibfnamefont {Z.~J.}\ \bibnamefont {Medin}}, \bibinfo
  {author} {\bibfnamefont {H.}~\bibnamefont {Park}}, \bibinfo {author}
  {\bibfnamefont {R.~M.}\ \bibnamefont {Rauenzhan}}, \bibinfo {author}
  {\bibfnamefont {H.~A.}\ \bibnamefont {Scott}}, \bibinfo {author}
  {\bibfnamefont {K.~S.}\ \bibnamefont {Anderson}}, \bibinfo {author}
  {\bibfnamefont {T.~J.~B.}\ \bibnamefont {Collins}}, \bibinfo {author}
  {\bibfnamefont {L.~M.}\ \bibnamefont {Green}}, \bibinfo {author}
  {\bibfnamefont {J.~A.}\ \bibnamefont {Marozas}}, \bibinfo {author}
  {\bibfnamefont {P.~W.}\ \bibnamefont {McKenty}}, \bibinfo {author}
  {\bibfnamefont {J.~H.}\ \bibnamefont {Peterson}}, \bibinfo {author}
  {\bibfnamefont {E.~L.}\ \bibnamefont {Vold}}, \bibinfo {author}
  {\bibfnamefont {C.}~\bibnamefont {Di~Stefano}}, \bibinfo {author}
  {\bibfnamefont {R.~S.}\ \bibnamefont {Lester}}, \bibinfo {author}
  {\bibfnamefont {J.~P.}\ \bibnamefont {Sauppe}}, \bibinfo {author}
  {\bibfnamefont {D.~J.}\ \bibnamefont {Stark}},\ and\ \bibinfo {author}
  {\bibfnamefont {J.}~\bibnamefont {Velechovsky}},\ }\href@noop {} {\bibfield
  {journal} {\bibinfo  {journal} {Physics of Plasmas}\ }\textbf {\bibinfo
  {volume} {29}},\ \bibinfo {pages} {083901} (\bibinfo {year}
  {2022})}\BibitemShut {NoStop}%
\bibitem [{\citenamefont {Higginson}\ \emph {et~al.}(2022)\citenamefont
  {Higginson}, \citenamefont {Strozzi}, \citenamefont {Bailey}, \citenamefont
  {MacLaren}, \citenamefont {Meezan}, \citenamefont {Wilks},\ and\
  \citenamefont {Zimmerman}}]{Higginson2022}%
  \BibitemOpen
  \bibfield  {author} {\bibinfo {author} {\bibfnamefont {D.~P.}\ \bibnamefont
  {Higginson}}, \bibinfo {author} {\bibfnamefont {D.~J.}\ \bibnamefont
  {Strozzi}}, \bibinfo {author} {\bibfnamefont {D.}~\bibnamefont {Bailey}},
  \bibinfo {author} {\bibfnamefont {S.~A.}\ \bibnamefont {MacLaren}}, \bibinfo
  {author} {\bibfnamefont {N.~B.}\ \bibnamefont {Meezan}}, \bibinfo {author}
  {\bibfnamefont {S.~C.}\ \bibnamefont {Wilks}},\ and\ \bibinfo {author}
  {\bibfnamefont {G.}~\bibnamefont {Zimmerman}},\ }\href@noop {} {\bibfield
  {journal} {\bibinfo  {journal} {Physics of Plasmas}\ }\textbf {\bibinfo
  {volume} {29}},\ \bibinfo {pages} {072714} (\bibinfo {year}
  {2022})}\BibitemShut {NoStop}%
\bibitem [{\citenamefont {Haines}\ \emph {et~al.}(2020)\citenamefont {Haines},
  \citenamefont {Keller}, \citenamefont {Marozas}, \citenamefont {McKenty},
  \citenamefont {Anderson}, \citenamefont {Collins}, \citenamefont {Dai},
  \citenamefont {Hall}, \citenamefont {Jones}, \citenamefont {McKay},
  \citenamefont {Rauenzahn},\ and\ \citenamefont {Woods}}]{Haines2020a}%
  \BibitemOpen
  \bibfield  {author} {\bibinfo {author} {\bibfnamefont {B.~M.}\ \bibnamefont
  {Haines}}, \bibinfo {author} {\bibfnamefont {D.~E.}\ \bibnamefont {Keller}},
  \bibinfo {author} {\bibfnamefont {J.~A.}\ \bibnamefont {Marozas}}, \bibinfo
  {author} {\bibfnamefont {P.~W.}\ \bibnamefont {McKenty}}, \bibinfo {author}
  {\bibfnamefont {K.~S.}\ \bibnamefont {Anderson}}, \bibinfo {author}
  {\bibfnamefont {T.~J.~B.}\ \bibnamefont {Collins}}, \bibinfo {author}
  {\bibfnamefont {W.~W.}\ \bibnamefont {Dai}}, \bibinfo {author} {\bibfnamefont
  {M.~L.}\ \bibnamefont {Hall}}, \bibinfo {author} {\bibfnamefont
  {S.}~\bibnamefont {Jones}}, \bibinfo {author} {\bibfnamefont {M.~D.~J.}\
  \bibnamefont {McKay}}, \bibinfo {author} {\bibfnamefont {R.~M.}\ \bibnamefont
  {Rauenzahn}},\ and\ \bibinfo {author} {\bibfnamefont {D.~N.}\ \bibnamefont
  {Woods}},\ }\href {https://doi.org/10.1016/j.compfluid.2020.104478}
  {\bibfield  {journal} {\bibinfo  {journal} {Computers and Fluids}\ }\textbf
  {\bibinfo {volume} {201}},\ \bibinfo {pages} {104478} (\bibinfo {year}
  {2020})}\BibitemShut {NoStop}%
\bibitem [{\citenamefont {Gittings}\ \emph {et~al.}(2008)\citenamefont
  {Gittings}, \citenamefont {Weaver}, \citenamefont {Clover}, \citenamefont
  {Betlach}, \citenamefont {Byrne}, \citenamefont {Coker}, \citenamefont
  {Dendy}, \citenamefont {Hueckstaedt}, \citenamefont {New}, \citenamefont
  {Oakes}, \citenamefont {Ranta},\ and\ \citenamefont {Stefan}}]{Gittings2008}%
  \BibitemOpen
  \bibfield  {author} {\bibinfo {author} {\bibfnamefont {M.}~\bibnamefont
  {Gittings}}, \bibinfo {author} {\bibfnamefont {R.}~\bibnamefont {Weaver}},
  \bibinfo {author} {\bibfnamefont {M.}~\bibnamefont {Clover}}, \bibinfo
  {author} {\bibfnamefont {T.}~\bibnamefont {Betlach}}, \bibinfo {author}
  {\bibfnamefont {N.}~\bibnamefont {Byrne}}, \bibinfo {author} {\bibfnamefont
  {R.}~\bibnamefont {Coker}}, \bibinfo {author} {\bibfnamefont
  {E.}~\bibnamefont {Dendy}}, \bibinfo {author} {\bibfnamefont
  {R.}~\bibnamefont {Hueckstaedt}}, \bibinfo {author} {\bibfnamefont
  {K.}~\bibnamefont {New}}, \bibinfo {author} {\bibfnamefont {W.~R.}\
  \bibnamefont {Oakes}}, \bibinfo {author} {\bibfnamefont {D.}~\bibnamefont
  {Ranta}},\ and\ \bibinfo {author} {\bibfnamefont {R.}~\bibnamefont
  {Stefan}},\ }\href@noop {} {\bibfield  {journal} {\bibinfo  {journal}
  {Computational Science \& Discovery}\ }\textbf {\bibinfo {volume} {1}},\
  \bibinfo {pages} {015005} (\bibinfo {year} {2008})}\BibitemShut {NoStop}%
\bibitem [{\citenamefont {Haines}\ \emph {et~al.}(2017)\citenamefont {Haines},
  \citenamefont {Aldrich}, \citenamefont {Campbell}, \citenamefont
  {Rauenzahn},\ and\ \citenamefont {Wingate}}]{Haines2017}%
  \BibitemOpen
  \bibfield  {author} {\bibinfo {author} {\bibfnamefont {B.~M.}\ \bibnamefont
  {Haines}}, \bibinfo {author} {\bibfnamefont {C.~H.}\ \bibnamefont {Aldrich}},
  \bibinfo {author} {\bibfnamefont {J.~M.}\ \bibnamefont {Campbell}}, \bibinfo
  {author} {\bibfnamefont {R.~M.}\ \bibnamefont {Rauenzahn}},\ and\ \bibinfo
  {author} {\bibfnamefont {C.~A.}\ \bibnamefont {Wingate}},\ }\href@noop {}
  {\bibfield  {journal} {\bibinfo  {journal} {Physics of Plasmas}\ }\textbf
  {\bibinfo {volume} {24}},\ \bibinfo {pages} {052701} (\bibinfo {year}
  {2017})}\BibitemShut {NoStop}%
\bibitem [{\citenamefont {Taitano}, \citenamefont {Chac{\'{o}}n},\ and\
  \citenamefont {Simakov}(2018)}]{Taitano2018}%
  \BibitemOpen
  \bibfield  {author} {\bibinfo {author} {\bibfnamefont {W.~T.}\ \bibnamefont
  {Taitano}}, \bibinfo {author} {\bibfnamefont {L.}~\bibnamefont
  {Chac{\'{o}}n}},\ and\ \bibinfo {author} {\bibfnamefont {A.~N.}\ \bibnamefont
  {Simakov}},\ }\href {https://doi.org/10.1016/j.jcp.2018.03.007} {\bibfield
  {journal} {\bibinfo  {journal} {Journal of Computational Physics}\ }\textbf
  {\bibinfo {volume} {365}},\ \bibinfo {pages} {173} (\bibinfo {year}
  {2018})}\BibitemShut {NoStop}%
\bibitem [{\citenamefont {Taitano}\ \emph
  {et~al.}(2021{\natexlab{a}})\citenamefont {Taitano}, \citenamefont {Keenan},
  \citenamefont {Chac{\'{o}}n}, \citenamefont {Anderson}, \citenamefont
  {Hammer},\ and\ \citenamefont {Simakov}}]{Taitano2021}%
  \BibitemOpen
  \bibfield  {author} {\bibinfo {author} {\bibfnamefont {W.~T.}\ \bibnamefont
  {Taitano}}, \bibinfo {author} {\bibfnamefont {B.~D.}\ \bibnamefont {Keenan}},
  \bibinfo {author} {\bibfnamefont {L.}~\bibnamefont {Chac{\'{o}}n}}, \bibinfo
  {author} {\bibfnamefont {S.~E.}\ \bibnamefont {Anderson}}, \bibinfo {author}
  {\bibfnamefont {H.~R.}\ \bibnamefont {Hammer}},\ and\ \bibinfo {author}
  {\bibfnamefont {A.~N.}\ \bibnamefont {Simakov}},\ }\href
  {https://doi.org/10.1016/j.cpc.2021.107861} {\bibfield  {journal} {\bibinfo
  {journal} {Computer Physics Communications}\ }\textbf {\bibinfo {volume}
  {263}},\ \bibinfo {pages} {107861} (\bibinfo {year}
  {2021}{\natexlab{a}})}\BibitemShut {NoStop}%
\bibitem [{\citenamefont {Taitano}\ \emph
  {et~al.}(2021{\natexlab{b}})\citenamefont {Taitano}, \citenamefont
  {Chac{\'{o}}n}, \citenamefont {Simakov},\ and\ \citenamefont
  {Anderson}}]{Taitano2021b}%
  \BibitemOpen
  \bibfield  {author} {\bibinfo {author} {\bibfnamefont {W.~T.}\ \bibnamefont
  {Taitano}}, \bibinfo {author} {\bibfnamefont {L.}~\bibnamefont
  {Chac{\'{o}}n}}, \bibinfo {author} {\bibfnamefont {A.~N.}\ \bibnamefont
  {Simakov}},\ and\ \bibinfo {author} {\bibfnamefont {S.~E.}\ \bibnamefont
  {Anderson}},\ }\href {https://doi.org/10.1016/j.cpc.2020.107547} {\bibfield
  {journal} {\bibinfo  {journal} {Computer Physics Communications}\ }\textbf
  {\bibinfo {volume} {258}},\ \bibinfo {pages} {107547} (\bibinfo {year}
  {2021}{\natexlab{b}})}\BibitemShut {NoStop}%
\bibitem [{\citenamefont {Froula}\ \emph {et~al.}(2006)\citenamefont {Froula},
  \citenamefont {Ross}, \citenamefont {Divol},\ and\ \citenamefont
  {Glenzer}}]{Froula2006}%
  \BibitemOpen
  \bibfield  {author} {\bibinfo {author} {\bibfnamefont {D.~H.}\ \bibnamefont
  {Froula}}, \bibinfo {author} {\bibfnamefont {J.~S.}\ \bibnamefont {Ross}},
  \bibinfo {author} {\bibfnamefont {L.}~\bibnamefont {Divol}},\ and\ \bibinfo
  {author} {\bibfnamefont {S.~H.}\ \bibnamefont {Glenzer}},\ }\href@noop {}
  {\bibfield  {journal} {\bibinfo  {journal} {Reviews of Scientific
  Instruments}\ }\textbf {\bibinfo {volume} {77}},\ \bibinfo {pages} {10E522}
  (\bibinfo {year} {2006})}\BibitemShut {NoStop}%
\bibitem [{\citenamefont {Henchen}\ \emph {et~al.}(2018)\citenamefont
  {Henchen}, \citenamefont {Sherlock}, \citenamefont {Rozmus}, \citenamefont
  {Katz}, \citenamefont {Cao}, \citenamefont {Palastro},\ and\ \citenamefont
  {Froula}}]{Henchen2018}%
  \BibitemOpen
  \bibfield  {author} {\bibinfo {author} {\bibfnamefont {R.~J.}\ \bibnamefont
  {Henchen}}, \bibinfo {author} {\bibfnamefont {M.}~\bibnamefont {Sherlock}},
  \bibinfo {author} {\bibfnamefont {W.}~\bibnamefont {Rozmus}}, \bibinfo
  {author} {\bibfnamefont {J.}~\bibnamefont {Katz}}, \bibinfo {author}
  {\bibfnamefont {D.}~\bibnamefont {Cao}}, \bibinfo {author} {\bibfnamefont
  {J.~P.}\ \bibnamefont {Palastro}},\ and\ \bibinfo {author} {\bibfnamefont
  {D.~H.}\ \bibnamefont {Froula}},\ }\href
  {https://doi.org/10.1103/PhysRevLett.121.125001} {\bibfield  {journal}
  {\bibinfo  {journal} {Physical Review Letters}\ }\textbf {\bibinfo {volume}
  {121}},\ \bibinfo {pages} {125001} (\bibinfo {year} {2018})}\BibitemShut
  {NoStop}%
\bibitem [{\citenamefont {Henchen}\ \emph {et~al.}(2019)\citenamefont
  {Henchen}, \citenamefont {Sherlock}, \citenamefont {Rozmus}, \citenamefont
  {Katz}, \citenamefont {Masson-Laborde}, \citenamefont {Cao}, \citenamefont
  {Palastro},\ and\ \citenamefont {Froula}}]{Henchen2019}%
  \BibitemOpen
  \bibfield  {author} {\bibinfo {author} {\bibfnamefont {R.~J.}\ \bibnamefont
  {Henchen}}, \bibinfo {author} {\bibfnamefont {M.}~\bibnamefont {Sherlock}},
  \bibinfo {author} {\bibfnamefont {W.}~\bibnamefont {Rozmus}}, \bibinfo
  {author} {\bibfnamefont {J.}~\bibnamefont {Katz}}, \bibinfo {author}
  {\bibfnamefont {P.~E.}\ \bibnamefont {Masson-Laborde}}, \bibinfo {author}
  {\bibfnamefont {D.}~\bibnamefont {Cao}}, \bibinfo {author} {\bibfnamefont
  {J.~P.}\ \bibnamefont {Palastro}},\ and\ \bibinfo {author} {\bibfnamefont
  {D.~H.}\ \bibnamefont {Froula}},\ }\href {https://doi.org/10.1063/1.5086753}
  {\bibfield  {journal} {\bibinfo  {journal} {Physics of Plasmas}\ }\textbf
  {\bibinfo {volume} {26}} (\bibinfo {year} {2019}),\
  10.1063/1.5086753}\BibitemShut {NoStop}%
\bibitem [{\citenamefont {Milder}\ \emph {et~al.}(2019)\citenamefont {Milder},
  \citenamefont {Ivancic}, \citenamefont {Palastro},\ and\ \citenamefont
  {Froula}}]{Milder2019}%
  \BibitemOpen
  \bibfield  {author} {\bibinfo {author} {\bibfnamefont {A.~L.}\ \bibnamefont
  {Milder}}, \bibinfo {author} {\bibfnamefont {S.~T.}\ \bibnamefont {Ivancic}},
  \bibinfo {author} {\bibfnamefont {J.~P.}\ \bibnamefont {Palastro}},\ and\
  \bibinfo {author} {\bibfnamefont {D.~H.}\ \bibnamefont {Froula}},\
  }\href@noop {} {\bibfield  {journal} {\bibinfo  {journal} {Physics of
  Plasmas}\ }\textbf {\bibinfo {volume} {26}},\ \bibinfo {pages} {022711}
  (\bibinfo {year} {2019})}\BibitemShut {NoStop}%
\bibitem [{\citenamefont {Foo}, \citenamefont {Schaeffer},\ and\ \citenamefont
  {Heuer}(2023)}]{Foo2023}%
  \BibitemOpen
  \bibfield  {author} {\bibinfo {author} {\bibfnamefont {B.~C.}\ \bibnamefont
  {Foo}}, \bibinfo {author} {\bibfnamefont {D.~B.}\ \bibnamefont {Schaeffer}},\
  and\ \bibinfo {author} {\bibfnamefont {P.~V.}\ \bibnamefont {Heuer}},\
  }\href@noop {} {\bibfield  {journal} {\bibinfo  {journal} {AIP Advances}\
  }\textbf {\bibinfo {volume} {13}},\ \bibinfo {pages} {115328} (\bibinfo
  {year} {2023})}\BibitemShut {NoStop}%
\bibitem [{\citenamefont {Froula}\ \emph {et~al.}(2011)\citenamefont {Froula},
  \citenamefont {Glenzer}, \citenamefont {Luhmann},\ and\ \citenamefont
  {Sheffield}}]{Froula2011}%
  \BibitemOpen
  \bibfield  {author} {\bibinfo {author} {\bibfnamefont {D.~H.}\ \bibnamefont
  {Froula}}, \bibinfo {author} {\bibfnamefont {S.~H.}\ \bibnamefont {Glenzer}},
  \bibinfo {author} {\bibfnamefont {N.~C.}\ \bibnamefont {Luhmann},
  \bibfnamefont {Jr.}},\ and\ \bibinfo {author} {\bibfnamefont
  {J.}~\bibnamefont {Sheffield}},\ }\href@noop {} {\emph {\bibinfo {title}
  {{Plasma Scattering of Electromagnetic Radiation}}}}\ (\bibinfo  {publisher}
  {Academic Press},\ \bibinfo {year} {2011})\BibitemShut {NoStop}%
\bibitem [{\citenamefont {Baalrud}(2012)}]{Baalrud2012}%
  \BibitemOpen
  \bibfield  {author} {\bibinfo {author} {\bibfnamefont {S.~D.}\ \bibnamefont
  {Baalrud}},\ }\href {https://doi.org/10.1063/1.3690093} {\bibfield  {journal}
  {\bibinfo  {journal} {Physics of Plasmas}\ } (\bibinfo {year} {2012}),\
  10.1063/1.3690093}\BibitemShut {NoStop}%
\bibitem [{\citenamefont {Henchen}(2018)}]{Henchen2018b}%
  \BibitemOpen
  \bibfield  {author} {\bibinfo {author} {\bibfnamefont {R.~J.}\ \bibnamefont
  {Henchen}},\ }\emph {\bibinfo {title} {{Direct Measurements of Nonlocal Heat
  Flux in Laser-Produced Coronal Plasmas Using Thomson Scattering from
  Electron-Plasma waves}}},\ \href@noop {} {\bibinfo {type} {Doctoral
  dissertation}},\ \bibinfo  {school} {University of Rochester} (\bibinfo
  {year} {2018})\BibitemShut {NoStop}%
\bibitem [{\citenamefont {Le~Pape}\ \emph {et~al.}(2019)\citenamefont
  {Le~Pape}, \citenamefont {Divol}, \citenamefont {Huser}, \citenamefont
  {Katz}, \citenamefont {Kemp}, \citenamefont {Ross}, \citenamefont {Wallace},\
  and\ \citenamefont {Wilks}}]{LePape2019}%
  \BibitemOpen
  \bibfield  {author} {\bibinfo {author} {\bibfnamefont {S.}~\bibnamefont
  {Le~Pape}}, \bibinfo {author} {\bibfnamefont {L.}~\bibnamefont {Divol}},
  \bibinfo {author} {\bibfnamefont {G.}~\bibnamefont {Huser}}, \bibinfo
  {author} {\bibfnamefont {J.}~\bibnamefont {Katz}}, \bibinfo {author}
  {\bibfnamefont {A.}~\bibnamefont {Kemp}}, \bibinfo {author} {\bibfnamefont
  {J.~S.}\ \bibnamefont {Ross}}, \bibinfo {author} {\bibfnamefont
  {S.}~\bibnamefont {Wallace}},\ and\ \bibinfo {author} {\bibfnamefont
  {S.}~\bibnamefont {Wilks}},\ }in\ \href@noop {} {\emph {\bibinfo {booktitle}
  {61st Annual Meeting of the APS Division of Plasma Physics}}}\ (\bibinfo
  {year} {2019})\BibitemShut {NoStop}%
\end{thebibliography}%

\end{document}